\documentclass[a4paper,draftcls,journal,onecolumn]{IEEEtran}
\usepackage[utf8]{inputenc}
\usepackage{tabularx}
\ifCLASSINFOpdf
	\usepackage[pdftex]{graphicx}
	\DeclareGraphicsExtensions{.pdf,.jpeg,.jpg,.png}
\else
	\usepackage[dvips]{graphicx}
	\DeclareGraphicsExtensions{.eps}
\fi

\title{Delay Impacts on EEG-Based Determination of the Human Visual Interface QoE for Virtual and Augmented Realities}
\author{Frank H. P. Fitzek and Patrick Seeling%
\thanks{F. H. P. Fitzek is with the Centre for Tactile Internet with Human-in-the-Loop and the Deutsche Telekom Chair of
	Communication Networks, Technische Universität Dresden, Dresden, Saxony, Germany, email: frank.fitzek@tu-dresden.de}%
\thanks{Patrick Seeling is with the  Department of Computer Science at Central Michigan University, Mount Pleasant, MI, USA, email: patrick.seeling@cmu.edu}%
\thanks{Funded by the German Research Foundation (DFG, Deutsche Forschungsgemeinschaft) as part of Germany’s Excellence Strategy – EXC 2050/1 – Project ID 390696704 – Cluster of Excellence “Centre for Tactile Internet with Human-in-the-Loop” (CeTI) of Technische Universität Dresden.
}
}

\title{Why We Should NOT Talk about 6G}

\begin{document}
\maketitle
\begin{abstract}
While 5G mobile communication systems are currently in deployment, researchers around the world have already started to discuss 6G technology and funding agencies started their first programs with a 6G label.  Although it may seem like a good idea from a historical point of view with returning generations every decade, this contribution will show that there is a great risk of introducing 6G labels at this time.  While the reasons to not talk about 6G yet are manifold, some of the more dominant ones are $i.)$ there exists a lack of real technology advancements introduced by a potential 6G system; $ii.)$ the flexibility of the 5G communication system introduced by softwarization concepts, such as in the Internet community, allows for daily updates; and $iii.)$ introducing widespread 6G discissions can have a negative impact on the deployment and evolution of 5G with completely new business cases and customer ecosystems compared to its predecessors. Finally, as we do not believe that 5G is the end of our journey, we will provide an outlook on the future of mobile communication systems, independent of the current mainstream discussion.
\end{abstract}


\section{Introduction -- The Evolution of Mobile Communication Systems}
\label{s:intro}
Whether there is a clear need for a sixth generation (6G) of mobile communication systems can only be answered through an initial recognition of the history of mobile communications and the motivations underpinning the continuous development efforts.  Mobile communication systems are commonly categorized in generations. (We refer to mobile communication systems throughout this contribution with a focus on \textit{cellular} mobile communication systems.)  So far, a new generation has been introduced  in each decade. The first generation (1G) of mobile communication systems allowed a small group of privileged users to experience mobile voice services employing purely analog technology~\cite{Frenkiel10}. The second generation (2G) democratized these services for the masses exploiting digital technology. Both generations have been straightforward extensions of the existing public switched telephone networks (PSTNs) of their times to mobile services. The main services of 2G, following the global system for mobile communications (GSM) definition, were speech services, mobility, and security.  Message and data service implementations were only added to the network afterwards, featuring limited data rates and capacities~\cite{Rap02:Survey}. With the introduction of the third generation (3G), an extension of mobile communication systems to the Internet was targeted. Even though the first realizations, such as i-mode in Japan or the wireless application protocol (WAP) in Europe, had been great failures, the fourth generation (4G) finally enabled the mobile Internet for users with great success~\cite{Bjerke11}.
The introduction of a new generation was always motivated by and aligned with changes in the underpinning technical details. While 1G was based purely on analog technology,  the digital era was introduced with 2G. Furthermore, 2G in Europe was based on time division multiple access (TDMA), while code division multiple access (CDMA) was deployed in the USA and other countries alongside TDMA. With 3G, CDMA was introduced as a global standard, significantly unifying the formerly divided technology landscape. Finally, 4G uses orthogonal division multiple access (OFDM). With that, each generation could have been identified by the predominant access scheme. All generations so far have in common that they targeted the consumer market’s main demands over time: to enable voice services and mobility in the beginning (starting in the 1980s) up to video streaming and social networking towards the end (the 2010s). Nevertheless, each generation has its entitlement in a technical development stage.
The fifth generation (5G) is different~\cite{Molisch5G17}, as $i.)$ it continues to use OFDM as it provides technical benefits over other access technologies and there are simply no novel access technologies left, $ii.)$ its standardization efforts from the beginning onward did not just follow a philosophy of pure capacity increases but already addressed new technical parameters, such as latency, resilience, heterogeneity, or energy, $iii.)$ it already has a strong impact on the overall network architecture rather than focusing on a new air interface with some radio access network (RAN) elements, and $iv.)$ the target use cases are mainly in the domain of machine control rather than the consumer market, a major shift with respect to prior generations.

\begin{figure}[htbp]
    \centering
    \includegraphics[width=\textwidth]{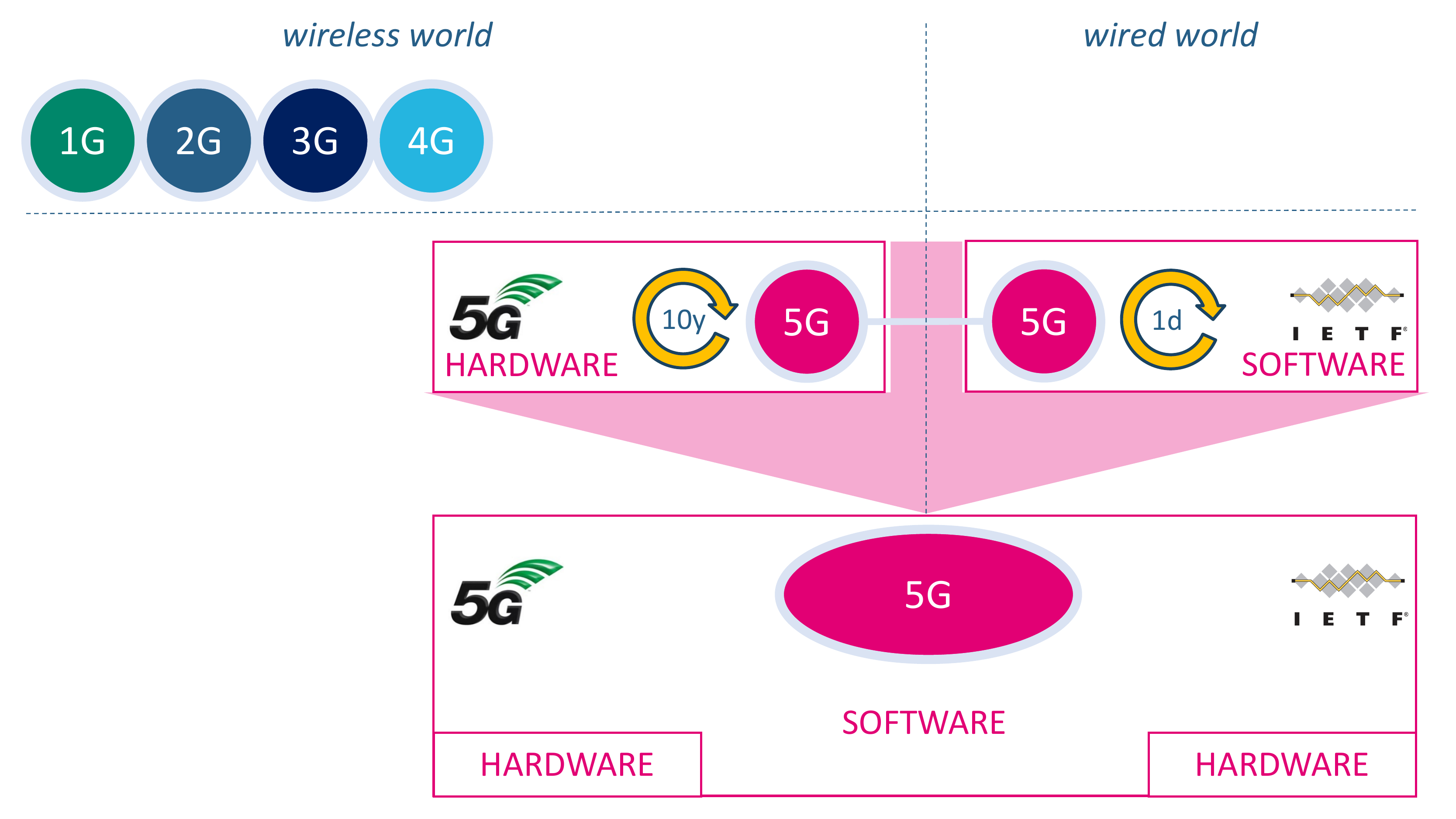}
    \caption{Evolution of mobile communication generations.}
    \label{fig:fig1}
\end{figure}

As illustrated in Figure~\ref{fig:fig1}, the initial four generations have been addressing the wireless domain with some additional RAN elements. 5G will follow this approach on the wireless side, but is strongly engaged with the wired domain. One evidence for that is the large number of bilateral meetings from the 3rd Generation Partnership Project (3GPP) and the Internet Engineering Task Force (IETF) to define 5G technologies. These meetings are embossed by a cultural clash where one side is dominated by hardware while the other is dominated by software.  While the generations on the mobile communication system side are a result of changing hardware in the time frame of one decade, the Internet domain is realized by software and allows for daily updates.  Once the two worlds are truly merged in an holistic manner in the future, software will play a dominant role, while specialized hardware is needed only for special tasks, such as energy efficient computing or energy efficient and secure radio front-ends with antennas.

5G is a highly disruptive communication system, as can already be derived from only some of the main differences to its predecessors, as highlighted in the following.
\begin{description}
\item[Low Latency as a Service:]~\\Besides pure data rate, 5G also targets low latency communications.  This enables communication for time-critical scenarios. In other words, 5G combines control and communication theory. While the predecessors target the consumer market for communication needs of humans, 5G is mainly targeting machine-type communications. Even though 5G is currently introduced as a bit pipe, with upcoming releases 16 and onward, the new service area will be the target. Therefore, 5G is able to become the communication network for a broad variety of use cases, such as transportation, Industry~4.0, Agriculture~4.0, construction, health care, or the upcoming Tactile Internet. The Tactile Internet has the most severe latency requirements with $1$~ms end-to-end delay including the embedded system, the wireless link, and the computing within the wired network. As given in~\cite{Xian1803:Reducing}, the most challenging entity for low latency are network functions realized in software.

\item[Softwarization:]~\\5G is not just an agnostic bit pipe as the generations before. This is achieved by the concept of softwarization. While its predecessors came in the form of dedicated hardware, 5G is mainly based on generic hardware  with software solutions. Different concepts, such as software defined radio (SDR), software defined networks (SDN), and network function virtualization (NFV), enable network operators to keep operational expenses (OPEX) as well as capital expenses (CAPEX) low. Additionally, softwarization enables network operators to dynamically reorganise their networks for different current and future use cases. New concepts such as mobile edge cloud and network slicing are only possible due to the introduction of the softwarization concept.

\item[5G as Campus Solutions:]~\\Due to the last point, the realization of so-called 5G campus solutions is now possible. In some countries, industry or government entities can apply for 5G frequencies for a given local area. A 5G campus solution operates in an isolated geographic area and provides localized services. Mobility support is not needed in this application scenario, but ultra-low latency communication is.
\end{description}

These three main differentiators of 5G with respect to prior generations already highlight how it completely differs from its predecessors. It is the first generation to target machine-type communications, even though the support of multimedia services is still an option. 5G will support different technical parameters such as low latency, security, heterogeneity, and more. More importantly, the replacement of specialized hardware with generic hardware in combination with dedicated software solutions is such a fundamental change, that it will have long-term implications for the future of communication networks.

Based on the paradigm shift towards network as a software solution, the architecture of 5G will be extended by campus solutions. Those campus solutions do not necessarily support mobility in the cellular WAN context, but rather offer mobility in the LAN context, which is more nomadic.
While earlier generations defined communication as a transport of bits leading to the end-to-end paradigm of communication, with 5G we enter a new definition where communication is defined as transport, storage, and computing of bits within the network. The disruptive nature of 5G is that it elevates the network from data to information.

Despite these revolutionary changes, it is highly likely that 5G will face immense problems in the beginning of its deployment phase.  Classic services, such as video streaming, voice calls, etc., are part of the general update of capabilities from 4G to 5G -- but focusing on their contribution to business success might be more than disappointing. This should not surprise, as the  superiority of 5G lies in low latency communications, which will become part of implementation rollouts with the standard release 16 and onward.  Once the implmentations find their way into associated use cases, the business side of the success story of 5G will start as well.

\section{The Atom -- A Holistic View on 5G}
\label{s:atom}

\begin{figure*}[htbp]
    \centering
    \includegraphics[width=0.9\linewidth]{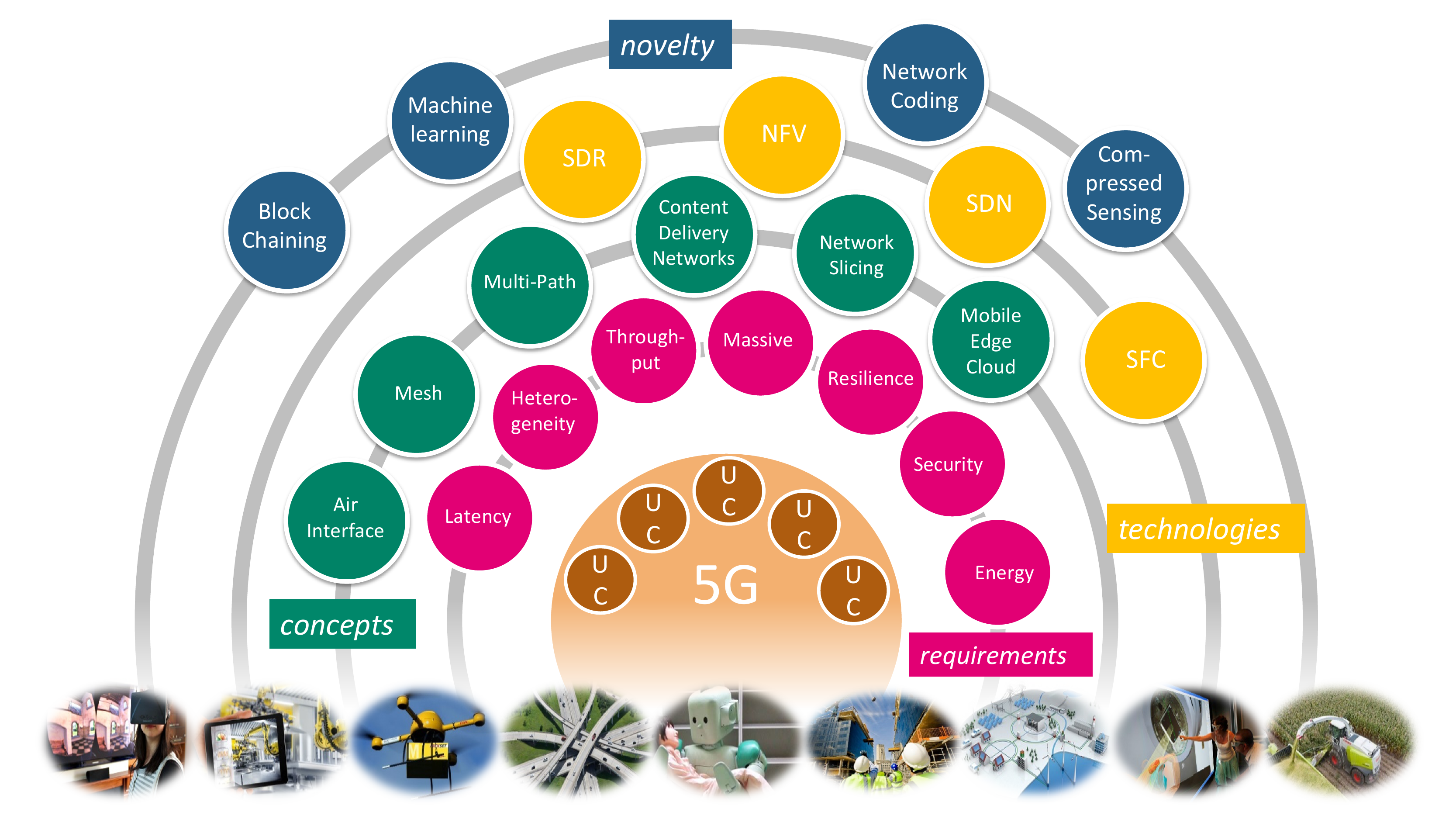}
    \caption{The 5G Atom -- A holistic view of 5G use cases, requirements, technologies, concepts, and novel applications.}
    \label{fig:atom}
\end{figure*}

In this section, we briefly discuss the 5G landscape from a holistic viewpoint by employing the \textit{5G Atom}.  Figure~\ref{fig:atom} illustrates this view, which deviates from the \textit{massiveness} in the International Mobile Telecommunications (IMT) vision~\cite{IMT2020} of massive IoT, massive multimedia, and massive low latency, by following 5G use cases defining higher layers, up to novel applications.

The \textit{massive} use cases identified in IMT-2020 are part of the initial wave of 5G use cases, but they do not account for all of them.  One can readily separate the use cases by their focus of communication, whereby the machine-to-machine communication will be expressed in roll-outs in connected autonomous cars, Industry~4.0, Agriculture~4.0, energy grids, as well as construction, health care, or education -- in what is heavily marketed today as the Internet of Things (IoT). In contrast to these, human-machine-type communications that are summarized as the Tactile Internet~\cite{FettweisTI} will feature ultra-reliable low latency communications (URLLC)~\cite{URLLC} enabling human-machine collaborations across all sectors of industry and in mixed realities.  The Tactile Internet has the potential to usher in a new paradigm shift from a network of information to a network of skills~\cite{CeTIweb} with a focus on layer 8 of the network.

A readily notable technical requirement resulting from these different use cases is the handling of increased heterogeneity of communication types and challenges when considering different communication actors and device types.  For example, connecting massive numbers of IoT devices in the billions requires new approaches to data handling and storage.  Additionally, 5G networks need to feature security as well as resilience, not just with respect to adverse actors threatening the network, privacy concerns, or interference.  Maintaining quality of service (QoS) levels to higher degrees than before for at least a slice of the network is of utmost importance as use cases feature control loops that are comprised of machines as well as humans, with potentially dire consequences as results of service degradation.  For example, URLLC requires significant reductions of the latency present in today's networks down to a millisecond, while simultaneously bounding delay variations (jitter).  As the case with all prior generations, increases of throughput are common requests in use case evolutions.  Some of these move from today's dominant multimedia streaming services to fully immersive experience streaming services requiring significantly more data.  With such a broad landscape of devices and use cases, energy consumption is an additional factor, e.g., to reduce battery depletion rates and required maintenance for devices in the field.

The following concept layer in Figure~\ref{fig:atom} highlights approaches to resolve these often diametrically opposed technical requirements.  The 5G new radio air interface will significantly reduce latency, offer higher throughput, and enable massive numbers of securely connected devices. In addition, new concepts that offer extensions of the 5G network through meshed networks as well as multi-path communications can aid significantly to realize the technical requirements as part of a softwarized solution.  In addition to networking-centric approaches, the softwarization inherent in 5G also is tied to new approaches to managing the network, such as focusing on information rather than just data, and to dynamically allocate the network resources (including computation and storage) through network slicing.  This approach enables an ability to dynamically move these resources at the network's edge, as mobile edge cloud (MEC).

These solutions make heavy use of softwarization technologies across the 5G network by employing software-defined radio (SDR), software-defined networking (SDN), network function virtualization (NFV), and service function chaining (SFC). Jointly, these technologies allow a dynamic allocation of all resources within a 5G network and their logical combination. This makes the common computing resources (communication, computation, and storage) flexible and versatile to perform flexible service composition.

The developed services can readily be altered to reflect changing use cases as well as new use cases providing network operators with long-term flexibility as well.  The outer layer application examples in Figure~\ref{fig:atom} that will likely become part of these future services include block chains, machine learning, network coding, and compressed sensing.  While  a detailed discussion of these are out of scope here, we note that all of these have been extensively researched in conjunction with networked applications.  It is therefore foreseeable that they will become part of future service offerings -- or not, as the flexibility offered by the softwarized network can readily support others not even considered yet.

\section{Mythbusting 6G}
\label{s:mb6g}

After the introduction of 5G, in this section, we present several theses which question the necessity of a 6G technology.

\subsection{Nothing New in 6G}
Currently, there is nothing to little novelty proposed for 6G that can not be adopted or realized in 5G networks and its evolution plan. Often new frequencies, such as terahertz frequencies, are labeled as 6G. But the 5G evolution foresees already higher frequencies up to 60\,GHz and to include even higher frequencies would not require a new generation. Even a change of hardware or software components is not necessarily leading to a new generation -- 2G already supported channel bundling (software) and novel modulation schemes (hardware). Some research labs are even proposing artificial intelligence or mobile edge cloud as a 6G feature, but this has been already introduced in 5G. Oftentimes, lower latencies than promised in 5G are also referred to when discussing 6G. Some researchers claim that higher data rates are needed to achieve this goal, others rely on a novel medium access protocol in their claims. In general, the data rates of 5G are already so high that the propagation delay increases in importance when compared to the relative delay, as~\cite{gallager} have shown for simple M/M/1 queues. In conclusion, there is little in individual technologies that would require a new generation. General evolution steps as well as disruptive ideas can be realized with software changes within 5G networks, as introduced beforehand.

\subsection{The Internet Has No Generation, Why Mobile Networks?}
Given the evolutionary changes in the roll-outs of cellular network generations to date, a common thread is the combination of typically specialized hardware combined with specialized and provider- or operator-based services targeting cellular end-systems used by humans. This development increased capacities on the network side to support mobile Internet based over-the-top (OTT) services at the end-units (e.g., smartphones for media consumption). The fixation on continuously evolving services for individual human customers meant that architectural changes were required  to fully enable Internet-based services following the end-to-end paradigm.

At the same time as mobile communication network operators switched out generations of different custom hardware and software solutions moving from 3G to 4G, the smartphones they were connecting to the Internet were already softwarized. The software-oriented approach inherent to common smartphones enables the continuous addition of new features based on a common set of hardware components via software updates.

Throughout the history of the Internet, the closest notion of generational change can be seen in the versions of protocols, predominantly the migration from IPv4 to IPv6. The reason that the sunset of IPv4 takes so long is also based on hardware legacy. Nevertheless, researchers in standardization groups such as IETF would never use the term generation as software enables agile testing and implementation. The overall nature of “Everything over IP and IP over Everything” abstracts or even virtualizes the underlying complexity for upper layer protocols. In turn, changes can be made on all layers above IP as needed in a fluid fashion.The softwarized nature of even a significant portion of the core network components means that within reason, new features can now be added via system software upgrades on general-purpose hardware and no longer require full system hardware swaps, i.e., a new generation.

As the network role is elevated from data to information, it becomes clear that only individual components need changes over time, as 5G already contains the built-in flexibility of softwarization of the network. Thus, it is difficult to justify individual component tweaks to modify network capabilities to be labeled as a generational change. Indeed, intuitive examples such as replacing an older general-purpose server in a rack with a more powerful could increase the in-network computation speed (e.g., to support more services offered through a MEC or in a nomadic 5G campus container), but hardly qualify as a new generation. Similarly, significantly changing TCP’s congestion algorithm did not result in new protocol versions despite significant performance changes. The network softwarization of 5G enables continuous, fluid change that reduces the need for generational changes.

\subsection{5G Engaged Vertical Players -- 6G Scares Them Away}
5G was very successful to engage vertical players even before the start of the technology itself. While 3G failed in identifying possible markets and 4G took over the promise of 3G, 5G breaks new grounds for a wide range of heterogeneous markets. Some markets such as agriculture or construction happily engaged with the 5G technology. Other market segments such as connected cars or Industry~4.0 had alternatives such as IEEE~802.11p or field busses. It took, and still takes, some effort to explain the advantages of 5G over existing solutions. This was successful for most use cases and is ongoing for others.
The upcoming discussions of the 6G technology creates some nervousness among the management in industry. When considered by merit and potential, there should be a Fear Of Missing Out (FOMO), but these discussions more likely begin to create a Fear Of Joining In (FOJI). Some might think that 5G is already an old technology and 6G is the holy grail now. This has the potential to destroy the momentum 5G has. Distracting the industry from 5G could lead to a decreased market adoption.  Subsequently, this would put both network operators and network equipment suppliers in great financial distress. Without the revenues earned  from 5G, a new generation, possibly 6G, would not even see the light of day.

\subsection{Only Even Numbered Generations Are a Success}
In several keynotes about 5G and 6G, the speakers claim that only even numbered generations are a success. Such statements are not only ridiculous but also technically wrong. These statements are  based on the difficulties 3G had in the beginning of its introduction, while 2G and 4G were a success story from the beginning. However, this argumentation has several flaws. The first generation 1G was undoubtedly a success even if not available for the masses. Furthermore, the problem with such a statement is the statistical confidence one can achieve with only five samples. Finally, if the statement were true, we should just use even numbered generations from the beginning.

\subsection{The Use Case, Stupid}
Making a numbers-only argument additionally obfuscates a significant shortcoming in the overall interplay of 5G and 4G with respect to use cases. Through complete softwarization of the network, components can be changed dynamically and without a need to call for a new number. It is noteworthy to consider the scenarios for which these last two generations were designed: mobile Internet user-centered in 4G (over-the-top media consumption by humans) and expanding with (mobile) mission-critical with ultra-low latency in 5G (massive machine-type communications, including the Tactile Internet). Some use cases for 5G, such as the outlined 5G campus, offer the seamless integration of solutions into a locally managed 5G solution, but no mobility.
Jointly, these cases cover the currently imaginable scenarios for implementation fairly well, as listed in Table~\ref{tab:one}.

\begin{table*}
\centering
\caption{Minimum requirements (in terms of generations and use cases) for human- and machine-type communications. \label{tab:one}}
\begin{tabularx}{\textwidth}{|l|l|X|}\hline
Mobility target    & Human                     & Machine  \\\hline
Mobile              &3G/4G for browsing   & 5G for cars, drones (for both mobility is needed), energy grids (coverage is needed), also for the Tactile Internet \\
                    &4G/5G for over-the-top services      & \\
                    &5G for the Tactile Internet      & \\\hline
Nomadic/stationary  & WiFi                      & 5G campus, OpenRAN, also for the Tactile Internet\\
                    &5G, also for the Tactile Internet & \\\hline
\end{tabularx}
\end{table*}

Differently put, there are no current new scenarios that are not already covered through 5G or with a potential incorporation of non-cellular approaches. The remaining generational updates could be attributed to  radio access network upgrades. While additional arguments could be made that increased radio access network capabilities would yield better front-end latency characteristics, the question is how much is enough for use cases?
As usual when moving into business considerations, one also needs to carefully consider the diminishing rates of return for increased resource expenditures. Realizing higher throughput to reduce transmission delays (assuming that propagation delays cannot be influenced in the near future, e.g., with quantum networking), higher frequency bands are needed. As a direct downside, this requires significant attenuation handling and will reduce the current small cell sizes for going beyond mmWave bands even further, which might impose the need for a significant and cost-prohibitive increase in the number of cells needed for sufficient coverage.

\section{So, What \textit{IS} the Future of Mobile Communication Networks?}
\label{s:future}
\begin{figure*}[htbp]
    \centering
    \includegraphics[width=0.9\linewidth]{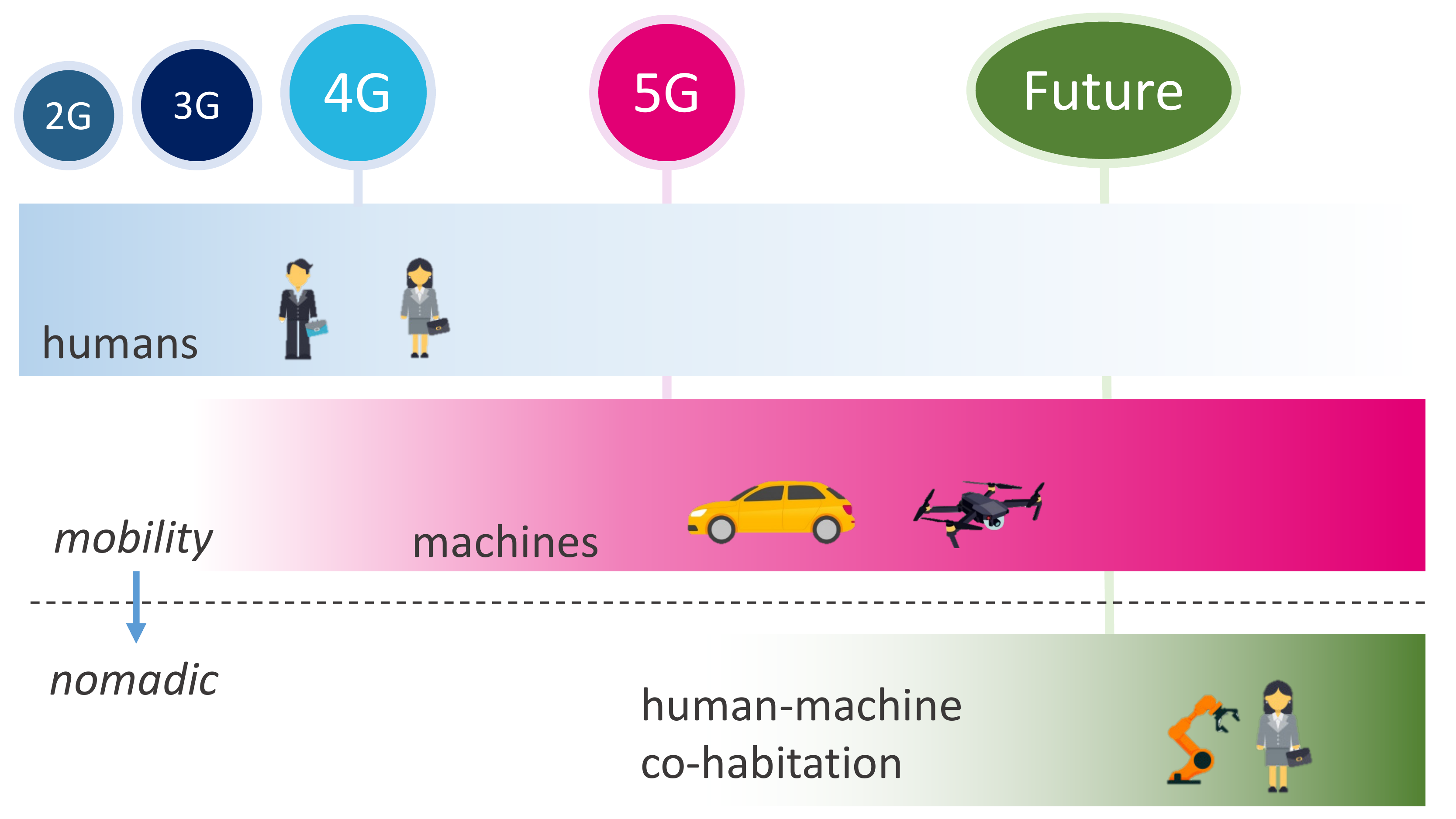}
    \caption{The evolution of mobile communication systems in generations and the disruptive change in future systems.}
    \label{fig:fig2}
\end{figure*}

While we advocate against the 6G label for the moment, we still believe in fundamental changes in future communication systems.
As illustrated in Figure~\ref{fig:fig2}, the first (cellular) mobile communication systems leading to 4G target humans and their mobility. Beginning with 5G, the focus shifts to being able to support communications with and between machines in real-time. As the 5G standard continues to evolve jointly with applied use cases, its evolution will continue with future releases 16, 17, \ldots, simultaneously incorporating additional features for deployments.

In contrast to the common prior focus on use cases incorporating mobility,  5G represents a turning point for the support of stationary, localized communications.  Similar to prior generations, 5G will continue to support the mobility of humans as well as machines, such as cars or drones.  The  URLLC components of 5G can even be seen as one enabler for future mobility scenarios with connected autonomous cars.  Mobility could also refer to the scope of information needs, which can readily go beyond single-cell sizes when considering the organization of large, distributed deployments, such as the energy grid.
While information mobility support is highly needed for these types of use cases, the emergence of human-machine co-habitation for the future of work and life represents a new scenario.  Here, URLLC will support the Tactile Internet-based and localized cooperation of humans and robots within their similarly networked but transparently operating local environment.

In these localized scenarios, mobility is just needed in a nomadic context common to WiFi networks, e.g., as provided by EduRoam for academic users that move between campuses, and the focus will be on URLLC. 5G will always be superior over WiFi (i.e., the IEEE 802.11 WLAN standard range) products, even WiFi6, when it comes to dedicated latency requirements. WiFi’s core philosophy of “listen before talk” clashes with deterministic latencies. Similarly, potential enhancements to WiFi protocols still require CSMA to operate in the general ISM bands, see, e.g.,~\cite{ETSI300338}, which will continue to result in a no-guarantee upper timing bound unsuitable for mission-critical application scenarios requiring URLLC.
Nevertheless, the cost efficiency of WiFi networks that maintain their ISM band reliance is still remarkable and readily serves as a target for future communication systems. This future might well be in the combination of the strengths of the two aforementioned technologies, namely 5G and WiFi6.

Subsequently, it is highly likely that we will witness two main trends, namely the evolution of 5G technologies and the rise of a new communication system that is not categorized in a generation. This new communication system will target specific use cases that do not need mobility support and national coverage to a large extent. Here we see Industry~4.0 or human--machine interaction as possible use cases. This can be seen as a low cost variant of the 5G campus idea, where the wireless technology is realized by OpenRAN solutions.

The flexibility of OpenRAN in combination with network slicing ideas of the core network will enable ultra-low latency communications that is also  resilient against hostile agents.  For example, adverse activities could include denial of service through extensive jamming of individual cells or modifying the data exchanged. The latter case is particularly interesting in the context of URLLC, as sophisticated mitigation techniques might readily impact the ability to remain at the ultra-low latency level, e.g., as required for the 1\,ms round-trip (including all sensing, computing, and actuating next to network transport) deadlines imposed by the Tactile Internet.  In the context of collaborative robotics in Industry~4.0 use cases, one result could be that adversaries become enabled to tamper with the actual payload without detection, e.g., through knowledge of the actual information content, for an individual robot’s control and cause undetectable extensive damage. One opportunity that OpenRAN offers with software-defined radios is to change the underlying MAC protocols at a high frequency. (We note that this is not the same approach as security through obfuscation, which is commonly considered harmful, but a new approach to security enabled by softwarization.) A similar trend to enable the lower latency bounds imposed by use cases is the need to move computation from the cloud to the edge of the network. Here, significant computational resources will need to be dynamically allocated and re-partitioned based on use cases. Again, this can only be realized through softwarization. Further latency reductions for actual use cases that take the shift to information-based mobile communications into account will, in turn, more likely be based on commodity hardware upgrades at the network edge and algorithmic improvements in software, as ultra-low latency dependent services will not be able to leave an operator’s network. Even more stringent latency requirements could be realized in 5G campus offerings with the possibility of significant embedded computational resources at its center (if an individual single cell) or edge (in general).

Considering the combination of flexible, commodity hardware paired with fully softwarized access and core networks as well as OpenRAN for the last hop, it becomes clear that 5G generates the ability for network operators to offer everything as a service (*aaS). For example, general latency as a service can be realized through flexible resource slicing on the network and computation side. Upgraded, guaranteed latencies can be provided through software-based OpenRAN changes that implement a bus-based polling protocol of subscribed stations, resulting in upper latency bounds. Storage, in-network computation, resource pooling, and, ultimately, services as a service (SaaS) all become enabled through elevation of the network to an information level, where an interplay of wireless access technologies could be orchestrated through 5G.

\section{Conclusion}
\label{s:conc}

In this article, we made a case for future mobile communication network research needs. Research and development must focus on the evolutionary advancement of the 5G roadmap. This roadmap provides many improvements in the areas of spectrum usage, integration of IETF software solutions for the mobile edge cloud, network slicing, and artificial intelligence. In parallel, there will be a disruptive evolution in the architectures of future mobile communications systems, targeting very specific applications, such as human--machine cooperation and Industry~4.0. The envisioned mobile radio systems will no longer be cellular and will only provide local coverage. The latter will allow low latency applications, novel security concepts, and extremely low cost. This approach could be seen as a symbiosis of the existing 5G campus solutions and WiFi6.
Nevertheless, we have also argued in this article that the use of the 6G label is not only counterproductive, but it also is not even technically motivated. Using 6G is counterproductive where new markets have just been opened up by 5G. The players in these new markets could be very nervous about investing in 5G, when new technology may soon be available. For example, the mere use of the term 6G may prevent the 5G market from fully developing. From a technical point of view, there is little reason to talk about a further generation in mobile communications because with 5G we leave the area of hardware behind us and now mostly realize everything in software. Thus, mobile communication systems are reaching the status that the Internet community has held for decades: continuous evolution and improvement through software. Nowadays, the 6G term is more a marketing tool that might end up in a 6G label without technical needs.


\end{document}